\documentclass[journal=jacsat,manuscript=article]{achemso}

\usepackage{chemformula} 
\usepackage[T1]{fontenc} 

\usepackage{amsmath}
\usepackage{amssymb}
\usepackage{float}
\usepackage{lineno}
\usepackage{wrapfig}
\usepackage{subcaption}
\usepackage{setspace}
\usepackage{epsfig}
\usepackage{amsthm}
\usepackage{lscape,graphicx}
\usepackage{array}
\usepackage{caption}
\usepackage{setspace}
\usepackage{fullpage}
\usepackage{centernot}
\usepackage{bbm}
\usepackage{chemarr}
\usepackage{changes}
\usepackage{tcolorbox}

\newcommand{\utwi}[1]{\mbox{\boldmath $ #1$}}

\newcommand{\bu}{{\utwi{u}}}

\newcommand{\bx}{{\utwi{x}}}

\newcommand{\bJ}{{\utwi{J}}}

\newbool{comments}
\booltrue{comments}

\providecommand{\keywords}[1]
{
  \small	
  \textbf{\textit{Keywords---}} #1
}

\author{Farid Manuchehrfar}

\author{Huiyu Li}

\author{Ao Ma}
\email{aoma@uic.edu}

\author{Jie Liang}
\email{jliang@uic.edu}

\affiliation{Center for Bioinformatics and Quantiative Biology and Richard and Loan Hill Department of Biomedical Engineering, University of Illinois at Chicago, Chicago, IL 60607.}

\title[Reactive Vortexes  in Activated Process]{\LARGE \bf
Reactive Vortexes in a Naturally Activated Process: Non-Diffusive  Rotational Fluxes at Transition State Uncovered by Persistent Homology}


\begin{document}

\newpage
\begin{abstract}
Dynamics of reaction coordinates during barrier-crossing are key to understand activated processes in complex systems such as proteins. The default assumption from Kramers’ physical intuition is that of a diffusion process. However, the dynamics of barrier-crossing in natural complex molecules are largely unexplored.   Here we investigate the transition dynamics of alanine-dipeptide isomerization, the simplest complex system with a large number of non-reaction coordinates that can serve as an adequate thermal bath feeding energy into the reaction coordinates.  We separate conformations along the time axis and construct the dynamic probability surface of reaction. We quantify its topological structure and rotational flux using persistent homology and differential form. Our results uncovered a region  with strong reactive vortex  in the configuration-time space, where the highest probability peak and the transition state ensemble are located. This reactive region contains strong rotational fluxes:   Most reactive trajectories swirl multiple times around this region in the subspace of the two most-important reaction coordinates. Furthermore, the rotational fluxes result from cooperative movement along the isocommitter surfaces and orthogonal barrier-crossing. Overall, our findings offer a first glimpse into the reactive vortex regions that characterize the non-diffusive dynamics of barrier-crossing of a naturally occurring activation process.

\vspace{.5in}

\end{abstract} \hspace{10pt}

\newpage
\section {Introduction}
Activated processes are ubiquitous in molecular systems, ranging from
 ring puckering in small molecules to
 protein conformational changes important for enzymatic catalysis.  The time-scale of activated processes is  characterized by their reaction rates.
Collectively, the proper timing of various activated processes give rise to proper functioning 
of cellular activities.

The conceptual foundation of the mechanism of activated processes is the transition state theory~\cite{wynne1935absolute,Wigner1938, Chandler1978,berne1988,Hanggi1990}. According to this theory, reaction occurs when a molecule crosses the energy barrier separating the reactant basin and the product basin.  
All reacting molecules must pass through the transition state, which are at the top of the energy barrier.  The method of committor test has been developed to 
assess rigorously whether a given molecular configuration is at the transition state~\cite{Du1998,Onsager1938,Bolhuis2002}. 
However, the transition state theory itself does not provide a dynamic model specifying how a molecule reaches the barrier top.
A number of models have been developed to  account for the dynamics of activation processes~\cite{berne1988,Hanggi1990}.  Following Kramer's theory~\cite{Kramers1940284},  the reaction process is often regarded as a particle moving along a 1-dimensional  double-well potential governed by Langevin equation~\cite{Kramers1940284}. 
 At the  small friction regime where molecular collisions change the overall energy of the system slowly and gradually, the total energy changes in a diffusive manner, and the system resides in the regime of energy diffusion.
 At the large friction regime where the total energy value fluctuates dramatically upon molecular collisions, one instead needs to identify the relevant reaction coordinates and model the barrier crossing events as a diffusion process along the reaction coordinates, leading to the spatial diffusion regime~\cite{Kramers1940284, zwanzig2001nonequilibrium}.  For proteins, the transition state theory  has been synthesized with the dynamics of spin-glasses, leading to insights into the process of protein folding~\cite{bryngelson1987spin,bryngelson1995funnels}. 
While modeling reaction dynamics with the underlying assumptions of a diffusion process has been fruitful, 
a basic question remains: the  details of the reaction dynamics of complex systems at the transition state are not known.  It is unclear if they are fully captured by diffusion models and if there are prominent non-diffusive behavior at the transition state.

A more direct approach to study
activated process is to investigate
molecular movements in the phase space of atomic positions and
momenta throughout the  process.
If the  bottlenecks to energy flow and the surface dividing the conformational basins satisfying the committor test
 can be located,
one can investigate the reaction rate  by studying the flux of phase points crossing the  bottleneck surfaces, also called the separatrix~\cite{davis1986unimolecular,zhao1992approximate,nagahata2021phase}.
For simple molecules where a
2-dimensional coordinate system is sufficient to describe both its
potential energy and kinetic energy, the bottleneck surfaces can be identified with accuracy.
A number of theoretical models of reaction rates have been developed based on analysis of the bottleneck surface and the crossing flux of phase space points for various small molecules~\cite{davis1986unimolecular,gray1987phase,marston1989reactive,de1989order,zhao1992approximate}.  
For example,
Zhao and Rice introduced a method to approximate the separatrix of 3-phospholene and studies the Poincare map of the dynamics systems governed by the equation of motion, from which the reaction rate was calculated from trajectory analysis\cite{zhao1992approximate}. 
Marston and De Leon developed a kinetic theory based on the discovery of the existence of reactive islands in the  3-phospholene system at a high energy level.  The reactive islands are found to be structured regions of reactivity in the phase space, from which pathways to reactions are determined\cite{de1989order,marston1989reactive}.
Other studies explored  the dividing surfaces and the structures of the phase space near regions corresponding to $k$-saddle points of the potential energy surface~\cite{collins2011index}.
A review on recent development in the analysis of geometrical structures of the 
phase space can be found in~\cite{nagahata2021phase}.
Nevertheless,  this direct approach is not feasible for molecules with a non-trivial number of degree of freedom, as it is not possible to construct the full energy surface and to explicitly define the dividing surfaces in high-dimension. Furthermore, the ability of generating full trajectories of the activated process may be very limited:  often only one or a few trajectories connecting the reactant and the product basins under natural  dynamics for proteins are accessible to simulations.  While reactive islands in the phase space has been discovered for the 3-phospholene molecule, their presence is found when the molecule is at a much higher overall energy state than the barrier
height~\cite{de1989order, marston1989reactive}, and the general applicability of reactive island is unknown.  Overall, whether the activated process at the transition state follows a diffusion process and whether higher order non-diffusive features exist has not been answered following this direct approach.

In this study, we investigate the activated process of alanine dipeptide isomerization  $C_{7eq}\rightarrow C_{7ax}$ in vacuum. Alanine dipeptide is a well studied model system that bears many hallmarks of more complex systems.  Its reaction coordinates have been well-characterized~\cite{Bolhuis2000,Ma2005Automatic,Li_Ma2016_Reaction_Mechanism,Li_Ma2020KineticEnergy,wu2022rigorous,wu2021mechanism}, and the large number of non-reaction coordinates serves as a thermal bath that feeds energy into the reaction coordinates.  In addition, a large number of reactive trajectories can be generated using transition path sampling~\cite{Chandler1978,Bolhuis2000,Ma2005Automatic}, so molecular flux in phase space can be studied in details.  Furthermore, recent development in data science methods of topological data analysis equipped us with tools to quantify the  topological structures of probability surface in high-dimensional
space~\cite{Edelsbrunner2002,carlsson2009topology, Manuchehrfar_Activated}, allowing accurate analysis of probability surface over complex configuration-time space. 

Here we examine the  dynamic probability surface of reactive molecules over the configuration-time space and analyze its topological structure. Instead of the free energy surface, we study the probability surfaces over the configuration space at different time points of the dynamic process. 
By adding time as the extra-dimension,
we align all reactive trajectories to the transition time as time zero through analysis of the one-dimensional reaction coordinate~\cite{wu2022rigorous}.  
Our results reveal the existence of a region of  reactive vortex at the transition state. First, instead of a saddle point of index-1,   probability mass of reactive molecular trajectories accumulates in a well-defined space-time region, and the transition state appears as a probability peak of critical index-0 instead of a saddle point.  Second, the probability flux exhibits strong and consistent vortex-like rotational pattern in this region.  These topological and dynamic structures of the probability surface at the transition state are beyond random fluctuations and cannot be modeled as a diffusion process.  Instead, our results suggest the existence of regions of reactive vortex with rich non-diffusive structures and significant  dynamic  rotational flux at the transition state.

\begin{figure}[!htbp]
    \begin{center}
     \includegraphics[width=0.33\textwidth]{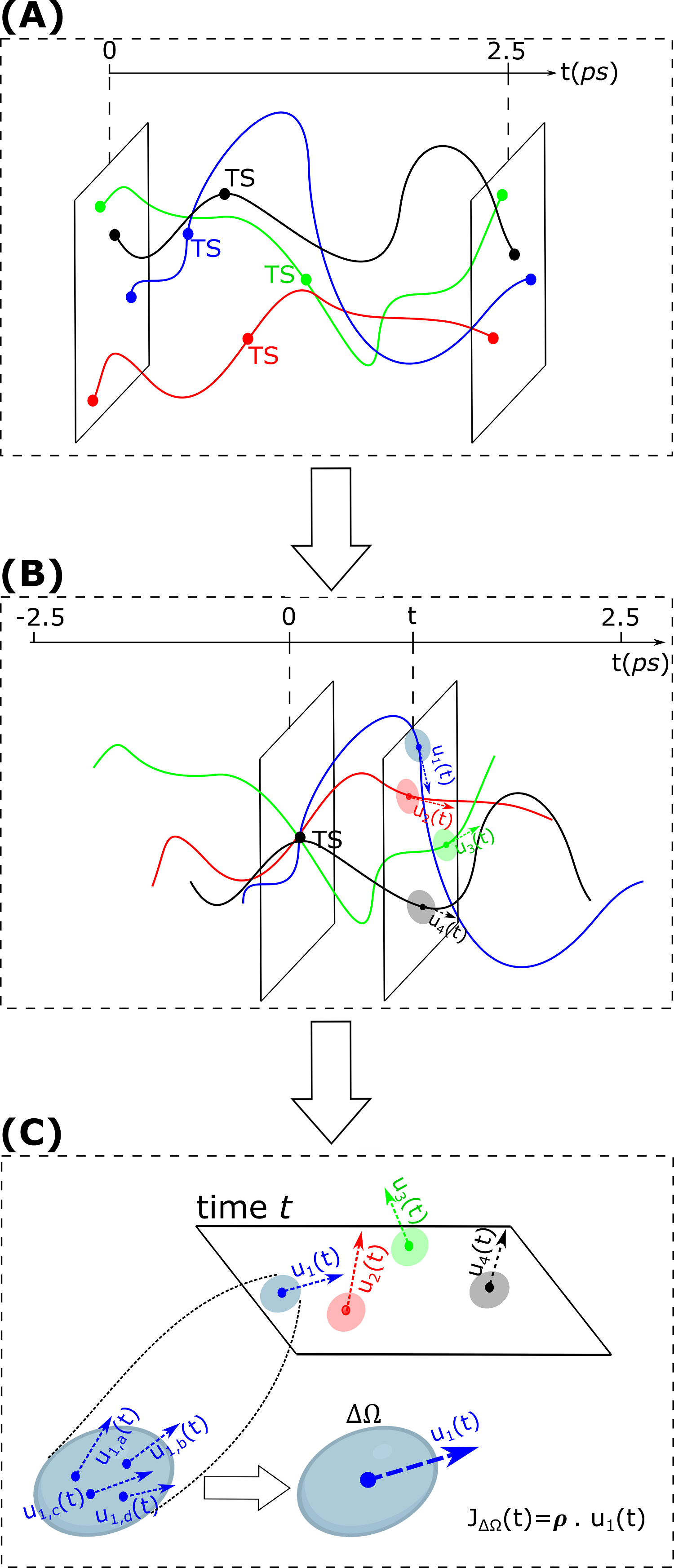}
    \end{center}
   \caption{\sf A schematic of time-alignment and trajectory flux.  \textbf{(A)} Samples of molecular dynamic trajectories along the axis of absolute time. The transition state occurs at different absolute times for different trajectories. \textbf{(B)} The trajectory time is off-set such that transition occurs at $t=0$ and all trajectories are aligned to  $t=0$. \textbf{(C)} After time-alignment, the fluxes of  trajectories at time $t$ over a neighborhood $\Delta\Omega$ is summed to calculated the total flux in the fixed volume $\Delta\Omega$.}    \label{fig:TPSIllustration}
\end{figure}

\begin{figure}[!htbp]
    \begin{center}
     \includegraphics[width=0.99\textwidth]{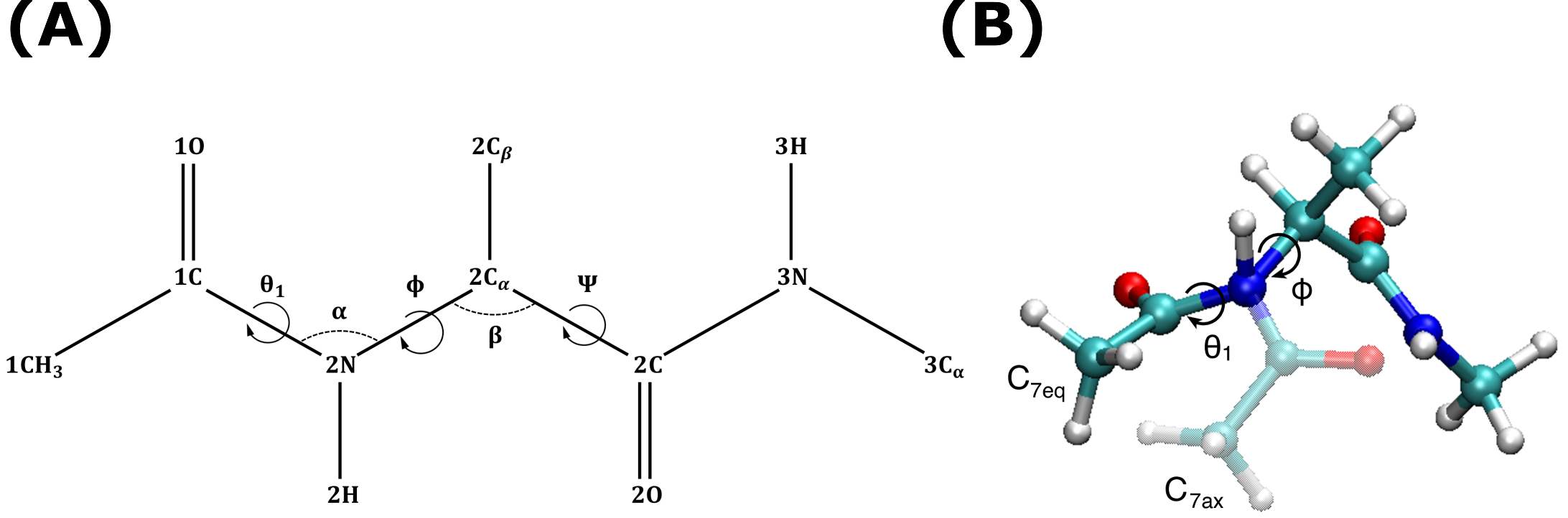}
    \end{center}
   \caption{\sf  The activation process of alanine dipeptide isomerization $C_{7eq}\rightarrow C_{7ax}$ in vacuum. (\textbf{A}) The five most important reaction coordinate revealed by the energy flow theory~\cite{Li_Ma2016_Reaction_Mechanism}.
   (\textbf{B}) $3$-d conformations before and after activation.}    \label{fig:alanine}
\end{figure}

\section {Theory, Models and Methods}
\subsection{Configuration Space, Time Alignment, and Dynamic Probability Surface}
\paragraph{Configuration space.}
We study the configuration space $\mathbb{M}_s$ of a molecule. 
A molecule may take a particular configuration $\bx  = (x_1,\, x_2, \cdots, x_d) \in \mathbb{M}_s$.
The configuration of alanine dipeptide is fully characterized by $d=60$ features~\cite{Li_Ma2016TPS}. 
  $\mathbb{M}_s$ therefore lies in a subspace of the Euclidean space $\mathbb{R}^d$: $\mathbb{M}_s \subset \mathbb{R}^d$, $d=60$.

\paragraph{Dynamic probability surface.}
We take the time $t$ as the $d+1$ dimension, and examine the  space-time relationship of the configuration of the molecule.
At time $t$, each configuration $(\bx;\; t)  = (x_1,\, x_2, \cdots, x_d;\; t) $ lies in the time-configuration space $\mathbb{M}_t \subset \mathbb{R}^{d+1}$, 
and 
has a probability $f(\bx;\; t) \in [0,1]$. Here the function
$
f: \mathbb{M}_t \rightarrow \mathbb{R}_{[0,\,1]}
$
 assigns the probability value $f(\bx; t)$ to a specific time-configuration $(\bx;\; t)$.  We study the topological structure of the dynamic probability surface $f(\bx; t)$ over $\mathbb{M}_t$, with time as the $(d+1)$-th dimension.

\paragraph{Time Alignment.}
The dynamic probability surface is constructed from sampled molecular dynamics trajectories.
All trajectories start from reactant basin and end in the product basin. Each trajectory is time-stamped by the duration from the start of the simulation, which is termed the {\it absolute time}.  Isomerization occurs in individual trajectory at different absolute time (Fig.~\ref{fig:TPSIllustration}A).
Since the relevant event is the transition of isomerization, 
we adjust the time with an offset so the transition time occurs at $t=0$ for each trajectory (Fig.~\ref{fig:TPSIllustration}B). Based on the 1-dimensional reaction coordinate computed following~\cite{wu2022rigorous}, we take the first time that the trajectory reaches the transition state as $t=0$.
While it is not possible to conduct committor test for each of the $3\times 10^6$ trajectories to determine the transition time, the occurrence of the isomerization is fully captured by the reaction coordinates.  Hence, we take the transition time $t=0$ as the time when the predicted committor value $p_B$ is 0.5 using the 1-D reaction coordinate described in the work done by Wu et al.~\cite{wu2022rigorous}, where the authors used a generalized work functional to summarize the mechanical effects of the couplings between different coordinates. Singular value decomposition (SVD) was then employed to extract the inherent structure of the generalized work function, from which the 1-dimensional reaction coordinate was  approximated with high accuracy.  This enabled identifications of the transition-state configurations where $p_B=0.5$ rapidly. More details can be found in~\cite{wu2022rigorous}.

\subsection{Flux and Its Rotation} \label{sec:flux}

\paragraph{Flux over the configuration space.} 
To characterize molecular movement in the configuration space,
we study its dynamic fluxes.
At time $t$, a molecule takes the configuration $\bx(t) \in \mathbb{M}_s \subset \mathbb{R}^d$ and has a velocity $\bu(t) \in \mathbb{R}^d$.

We first take the Lagrangian view, namely, the viewpoint of trajectories, where we start to  track the molecule at 
absolute
time $t'=0$ along the trajectory currently located at $\bx(0) \equiv \bx(t'=0)$ and float with this trajectory over time. 
The flux of this trajectory $f(\bx(0),t')$ at time $t'$ is then:
\begin{equation}
\bJ(\bx(0),\, t') =\rho\cdot\bu(\bx(0),\, t'),
\end{equation}
where $\rho$ is the weight of the trajectory and $\bu(\bx(0),\, t')$ is the velocity of trajectory at time $t'$.

We then take the Eulerian view and consider the fluxes associated with molecules located at specified locations. 
We consider a small fixed volume $\Delta \Omega \subset \mathbb{M}_s$ in the configuration space and measure the flux inside $\Delta \Omega$  at time $t$ 
after time alignment.
We do so by taking trajectories that are traveling inside $\Delta \Omega$ at time $t$.
The total flux in $\Delta \Omega$ at time $t$ 
is then:
\begin{equation}
\bJ_{\Delta \Omega}(t) 
= \int_{\bx(t) \in \Delta \Omega} \bJ(t) d \bx = \int_{\bx(t) \in \Delta \Omega} \rho\cdot\bu(\bx(0),\,t) d \bx ,
\end{equation}
where $\bx(t)$ is the location of the current flux line originate from $\bx(0)$, which has a velocity of $\bu(t)$ at time $t$ after alignment (Fig.~\ref{fig:TPSIllustration}C).

We estimate the fluxes from trajectories sampled by molecular dynamics. In this study, all trajectories are
properly generated without bias  and therefore of equal and constant weight proportional to $\rho$.
This is, as the MD trajectories are sampled without bias, 
$\rho$ is the same for all trajectories and is a constant 
over time.
The flux of the $i$-th trajectory at time $t$ is therefore:
\begin{equation}
\bJ_i(t) =\rho\cdot\bu_i(t),
\end{equation}
and the flux  $\bJ_{\Delta\Omega}(t)$ through a small volume $\Delta \Omega$ at time $t$ is:
\begin{equation}
\bJ_{\Delta\Omega}(t)
= \sum_{\bx_i(t) \in \Delta \Omega} \bJ_i(t) 
= \sum_{\bx_i(t) \in \Delta \Omega} \rho\cdot\bu_i(t).
\label{eqn:flux-disc}
\end{equation}
Here we set $\rho = 1/N$, where $N$ is the total number of unbiased trajectories.

\begin{figure}[!htbp]
    \begin{center}
     \includegraphics[width=0.99\textwidth]{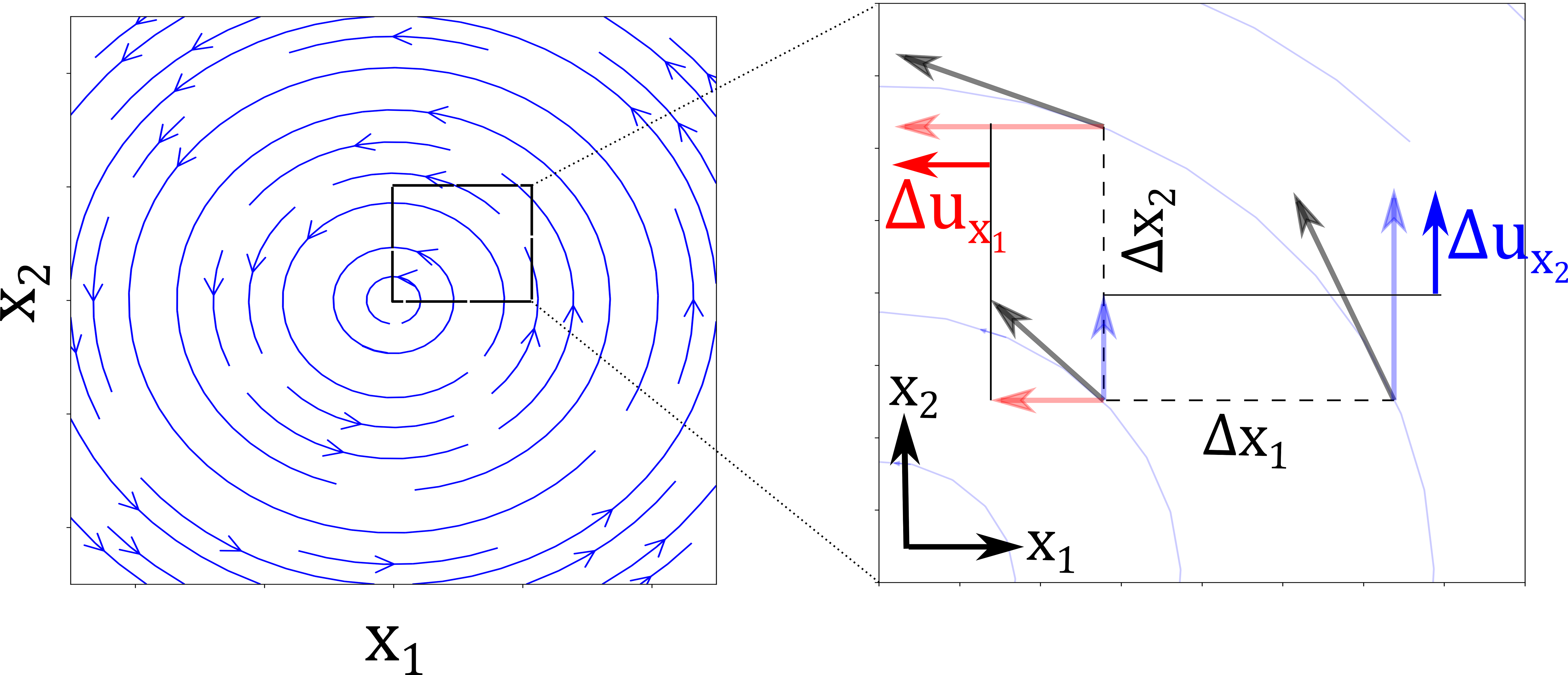}
    \end{center}
   \caption{\sf Rotational flux in a 2-dimensional velocity vector field, where $u_{x_1}=-x_2$ and $u_{x_2}=+x_1$. The velocity field circles  around the origin $(x_1,\, x_2)=(0,0)$ in the counter clockwise fashion. The rotation value  is the signed sum of changes of $u_{x_2}$ in the direction of $x_1$, and $u_{x_1}$ in the direction of $x_2$.}    \label{fig:CurlIllustration}
\end{figure}

\paragraph{Rotation of the flux.}
We further study the rotation of the flux. Our goal is to accurately characterize the activation dynamics during the barrier crossing process. Here we introduce a rigorous concept of rotational flux based on  differential form~\cite{bachman2012geometric} and describe a method for its computation. To illustrate,  let us examine a toy system of a velocity field over a 2-dimensional configuration space, where the velocity at each point $\bx=$($x_1,x_2$) is $\bu(\bx)=(u_{x_1},u_{x_2})= (-x_2, +x_1$). This velocity field exhibits a constant counter clockwise rotation around the origin %
(Fig.~\ref{fig:CurlIllustration}). The rotation of the velocity field on the $x_1$--$x_2$ plane is calculated by the difference of the changes of $u_{x_2}$ in the $x_1$ direction $\Delta u_{x_2}/\Delta x_1$  (blue, Fig.~\ref{fig:CurlIllustration}, right) 
where $\Delta u_{x_2} >0$ and $\Delta x_1 >0$, 
and changes in $u_{x_1}$ in the $x_2$ direction $\Delta u_{x_1}/\Delta x_2$  (red), where
$\Delta u_{x_1} <0$ and $\Delta x_2 >0$. 
Specifically, the rotation can be written as ($\Delta u_{x_2}/\Delta x_1 - \Delta u_{x_1}/\Delta x_2$).  

For flux in  high-dimensional space, we generalize the concept of rotation using differential form~\cite{bachman2012geometric}. 
The flux  $\bJ_{\Delta\Omega}(t)$ inside the small volume $\Delta \Omega$ is represented by a $d$-dimensional vector 
$\bJ_{\Delta\Omega}(t)=(J_{\Delta\Omega,\,1}(t), \cdots, J_{\Delta\Omega,\,d}(t)) \in \mathbb{R}^d$. 
This flux vector can be written as a $d$-dimensional $1$-form:
\begin{equation}
\bJ_{\Delta\Omega}(t)=J_{\Delta\Omega,\,1}(t) \cdot dx_1+ \cdots + J_{\Delta\Omega,\,d}(t) \cdot dx_d,
\end{equation}
where $J_{\Delta\Omega,\, i}(t) $ is the component of the flux in the $i$-th dimension.
The differential of this 1-form  can be written as 
$$d\bJ_{\Delta\Omega}(t)=
\sum_{\substack{i\neq j  \\i,j \in \{0,\cdots, d\}}}
(\frac{\partial J_{\Delta\Omega,\,j}(t)}{\partial x_i} - \frac{\partial J_{\Delta\Omega,\, i}(t)}{\partial x_j})  
\; dx_i \wedge dx_j$$ 
where $\wedge$ is the wedge operator denoting the exterior product~\cite{bachman2012geometric}. 
With the above definitions, the rotation of the $d$-dimensional flux vector can be written as 
\begin{equation}
\nabla \times J(\Omega,t)=<\frac{\partial J_{\Delta\Omega,j}(t)}{\partial x_i} - \frac{\partial J_{\Delta\Omega,i}(t)}{\partial x_j}>, \quad i\neq j \,\quad{\rm{ and }}\, \quad i,j \in \{0,\cdots, d\}
\label{eqn:rot}
\end{equation}
where 
$({\partial J_{\Delta\Omega,j}(t)}/{\partial x_i} - {\partial J_{\Delta\Omega,i}(t)}/{\partial x_j})$
represents the counter clockwise rotation of the flux projected onto the $i$-$j$ plane, as shown in Fig.~\ref{fig:CurlIllustration}.

\subsection{Topology of dynamic probability surfaces} 

\paragraph{Homology groups and persistent homology.} In this study, we investigate global features of the occurrence of probability peaks in the
time-configuration space. 
Our approach is that of homology group~\cite{munkres2018elements,hatcher2005algebraic} and persistent homology~\cite{Edelsbrunner:2230405,Edelsbrunner2002, carlsson2009topology}. 
Below we give a brief overview, as more detailed descriptions of this choice over that of critical points is described in ref~\cite{Manuchehrfar_Activated}.

Homology groups characterize holes of various dimension.  
Persistent homology quantifies the prominence of these holes.  Here we focus on the isolated probability peaks, which are components or 0-dimensional holes when they are isolated, and the configurations where they reside on.   
As an illustration, we envision that the probability landscape over the configuration space is flooded under the sea level. At the beginning, all mountain peaks on the probability landscape are below the sea level~(Supplementary movie 1). The sea level is then lowered gradually, with some peaks  emerging above the sea.  As the sea level further recedes, isolated mountain peaks may become connected by land-ridges.
Depending on how much of the time-space configurations have probability above a given  level, different peaks of probability over the configuration space may emerge.  As the  level is lowered, regions with probability greater than the given level enlarges.  As a result, previously isolated peaks may become land-connected and become merged into one connected component.

\begin{figure}[!htbp]
    \begin{center}
     \includegraphics[width=0.99\textwidth]{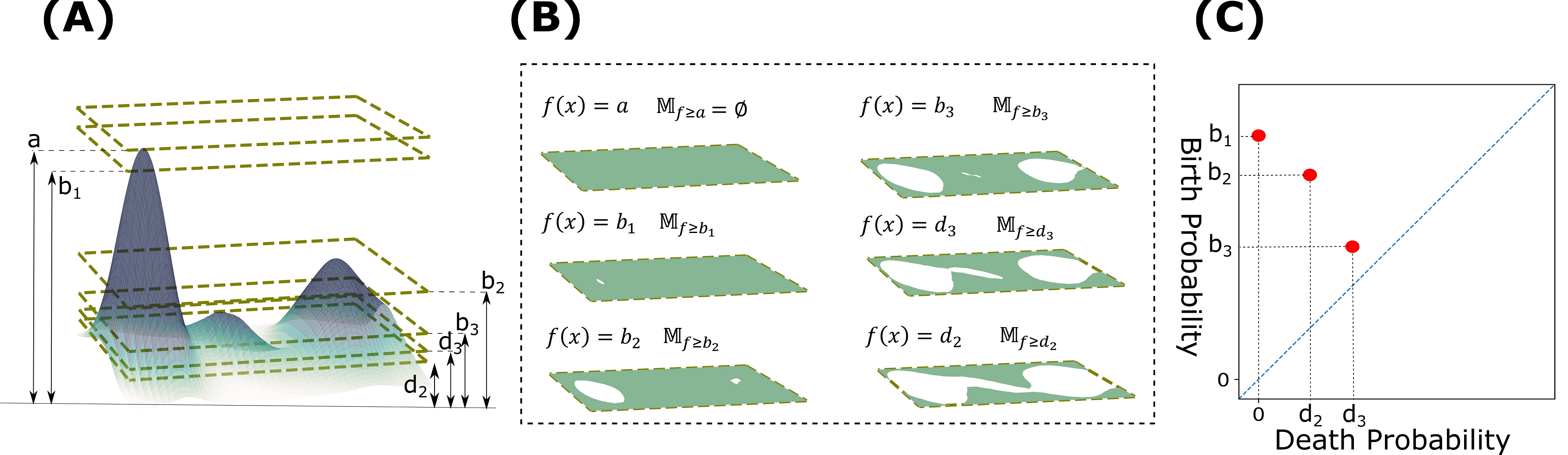}
    \end{center}
   \caption{\sf  The probability landscape $f(\mathbb{M})_t$ and the topology of its superlevel set $\mathbb{M}_{t,\,f\ge}$. 
   (\textbf{A}) The probability landscape and different sea levels. The superlevel set $\mathbb{M}_{t,\,f\ge .}$ are regions in $\mathbb{M}_t\subset \mathbb{R}^{(d+1)}$ whole probability height value is above the sealevel $(.)$.
   (\textbf{B}) At $f(\bx)=a$, the whole probability landscape is below the sea level and $\mathbb{M}_{t,\,f\ge a}=\emptyset$. 
   At $f(\bx)=b_1,\, b_2,\, b_3,\, d_3$, and $d_2$, the topology of the superlevel set changes.
   At each of $b_1$, $b_2$, and $b_3$, a new peak shown as a white island emerges. At $d_3$ and $d_3$, two separate peaks become merged together. 
   (\textbf{C}) The persistent diagram of the birth and death probabilities of the peaks. The sea levels of $b_1$, $b_2$, and $b_3$ are birth probabilities, and the sea levels of $d_3$ and $d_2$ are death probabilities.}    \label{fig:seaLevels}
\end{figure}

\paragraph{Superlevel sets and sublevel sets.}
Formally,  we can identify all $(\bx;\; t) \in \mathbb{M}_t$ with probability values 
$f(\bx;\; t)\ge a$. They form the
\textit{superlevel set} $\mathbb{M}_{t,\; f\ge a}$:
$$
\mathbb{M}_{t,\; f\ge a} \equiv \{ (\bx;\;t) \in \mathbb{M}_t| f(\bx;\;t) \ge a\} = f^{-1}([a,\,1)).$$
The \textit{sublevel sets} $\mathbb{M}_{t,\; f\le a}$ is defined similarly: 
$$
\mathbb{M}_{t,\; f\le a} \equiv \{ (\bx;\;t) \in \mathbb{M}_t| f(\bx) \le a\} = f^{-1}((0,\,a]).
$$

\paragraph{Time-space configurations as cubic complexes.}
We represent the $(d+1)$-dimensional time-space configuration space 
$\mathbb{M}_t$ using cubic complexes~\cite{kaczynski2006computational}. 
A $d$-dimensional cubic complex $K$ is the  union of points, line segments, squares, cubes, and their $k$-dimensional counterparts glued together properly, where $k \le d$ and all are of unit length, except points, which have no lengths.

\paragraph{Filtration.}
We examine the topological structures of probability peaks  on the time-configuration space,  and restrict ourselves to those  whose probabilities are above certain level.  By gradually adjusting this level, we can follow the details of topological changes.
As illustrated in Fig~\ref{fig:seaLevels}A-\ref{fig:seaLevels}B, the probability sea level at  $f(\bx) = a$ covers the whole probability landscape.  The domain of the portion of the landscape  above the sea level does not exist and is therefore the empty set $\emptyset$.
We gradually lower the sea level to value $b_1$, when the first peak emerges from the sea.
This sea level gives the birth of the first peak. 
At this time, we have the superlevel set $\mathbb{M}_{t,\; f\ge b_1}=\{\bx \in \mathbb{M}_t| f(\bx) \ge b_1 \}$, which are the set of configurations whose probability is 
$\ge~b_1$.
They form the small white region shown in Fig~\ref{fig:seaLevels}B (left, middle).  
We further lower the sea level to $b_2$  when 
the region associated with the first peak expands, and
another peak emerges above the sea. This is at the birth of the second peak (Fig~\ref{fig:seaLevels}B, left, bottom). At this time, we have the superlevel set $\mathbb{M}_{t,\;f\ge b_2}$. We then continue lowering the sea level to $b_3$, where a third peak emerges (Fig~\ref{fig:seaLevels}B, right, top). We have $\mathbb{M}_{t,\;f\ge b_3}$ at this level. 
We continue this process until sea level reaches $d_3$,  where the first and the third peaks are merged together by a land ridge  that has just emerged above the sea level. This is at the probability value of the death location of the third peak (Fig~\ref{fig:seaLevels}B, right, middle).  At this sea level, we have $\mathbb{M}_{t,\;f\ge d_3}$. We further decrease the sea level until we reach the sea level of $d_2$, when the second peak becomes merged with the other two peaks (Fig~\ref{fig:seaLevels}B, right, bottom). At this level, we have $\mathbb{M}_{t,\;f\ge d_2}$. 
At each of these levels, the topology of the superlevel set changes, namely, we have in sequence one component, two components, three components, two components, and then one component again.  These changes are captured by the changing 0-th homology groups or the Betti numbers  we  compute (see ref~\cite{Cohen-Steiner-stability,Manuchehrfar_Activated} for more details).

Formally, we have a descending sequence of probability values corresponding to the lowering sea level:
$$
1=a_0>a_1 > a_2 > \cdots > a_n = 0,
$$
and the corresponding superlevel sets, or the domains of the part of the landscape above the sea level, which are subspaces of $\mathbb{M}_t$:
$$
\emptyset =\mathbb{M}_{t,\;0 }
\subset \mathbb{M}_{t,\;1} 
\subset \mathbb{M}_{t,\;2} 
\cdots
\subset \mathbb{M}_{t,\;n} = \mathbb{M}_t. 
$$
As the full time-configuration space $\mathbb{M}_t$ is represented by a cubic complex $K$, each superlevel set $\mathbb{M}_{t,\;i}$ is represented by a subcomplex $K_i \subset K$, which can be derived from the original full complex $K$. The corresponding sequence of subcomplexes are:
$$
\emptyset = K_0
\subset K_1 \subset K_2 \cdots \subset K_n = K.
$$
This sequence of subcomplexes forms a \textit{filtration}.  

\paragraph{Persistence and persistent diagram.}
Upon changing the sea level so the corresponding subcomplex changes from $K_{i-1}$ to $K_i$, we may gain a new peak,
or we may lose one when a peak is merged with another one. A peak (or a connected component) is \textit{born} at $a_i$ if it is  present in  $K_i$ but absent in $K_{i-1}$ for any value of $a_{i-1} < a_i$.  The peak 
\textit{dies} at $a_i$  if it 
is present in $K_{i-1}$ but not at $a_i$ for any value of $a_{i-1} < a_i$.  
We record the location and the value of $a_i$,  namely, the corresponding $k$-cube and its probability value whose inclusion lead to the birth or the death event. 
 
The prominence of the topological feature of a peak is encoded in its life-time or \textit{persistence}. Denote the birth value  and the death value of peak $i$ as $b_i$ and $d_i$, respectively.
The \textit{persistence} of peak $i$ is $b_i-d_i$.
In the example shown in Fig.~\ref{fig:seaLevels}C, the component associated with the first, second and third peak is born at $f(\bx)=b_1$, $f(\bx)=b_2$, and $f(\bx)=b_3$, respectively.
At $f(\bx)=d_2$, the first and the second components are merged together. That is, the second peak  dies at $d_2$, and the persistence of this peak is $b_2-d_2$. At $f(\bx)=d_3$, the first and the third component are merged together, and the third peak  dies at $d_3$. The persistence of the third peak is therefore $b_3-d_3$.   The first peak dies at $f(\bx)=0$, and its persistence is $b_1-0=b_1$.

We record the birth and death events of  the peaks in a two-dimensional plot, or the \textit{persistent diagram}~\cite{Cohen-Steiner-stability}. Each peak is represented by a point in this diagram, where the birth value $b_i$ and the death value $d_i$  are taken as its coordinates ($b_i,\, d_i$). Fig.~\ref{fig:seaLevels}C shows the persistent diagram of our illustrative example.

\paragraph{Computation.}
We use the cubical complexes described in ref~\cite{wagner2012efficient} to calculate the persistent homology of the high dimensional time-evolving dynamic probability surface. The algorithm keeps track of changes in the super level set $\mathbb{M}_{t,\; f\ge a}$ of the probability surface, and considers  the birth and death of probability peaks.
  We neglect other topological properties such as 1-cycles. The locations $\bx_s$ where birth and death events occur, namely, the corresponding $k$-cubes are also computed. Details and code are  available at ``{https://github.com/fmanuc2/0-Homology-Group.git}''.

\section {Results}

\subsection{Model System and Computations.} 
\paragraph{Molecular Dynamics Simulation.}
Simulations were performed using the molecular dynamics software suite GROMACS-$4.5.4$~\cite{Berk2008}, with implementation of transition path sampling reported in ref.~\cite{Li_Ma2016TPS}.
Amber94 force field was used in our simulations~\cite{cornell1996second}. The  structure of the alanine dipeptide was energy minimized using the
steepest descent algorithm and heated to $300$K using velocity rescaling, with a coupling constant of
$0.3$ ps. The system was then equilibrated for $200$ ps. No constraints were applied. The time step
of integration was $1$ fs. We then performed $2$ ns NVE simulation, such that we are able to harvest one
reactive trajectory. The reactant basin $C_{7eq}$ was defined as $-3.49<\phi<-0.96$ and $-1.57<\psi<3.32$, and the product basin $C_{7ax}$ was defined as $0.87<\phi<1.74$ and $-1.39<\psi<0$~\cite{Li_Ma2016_Reaction_Mechanism}. Given this initial reactive path, $3 \times 10^6$ reactive trajectories were harvested through
transition path sampling.
Specifically, we randomly select one time point in the original reactive trajectory, exert a small
perturbation to the momentum, then initiate simulation from this point both forward and backward
in time. 
Simulations are performed with constant total energy of $36$KJ/mol, such that the average
temperature is 300 K in the transition path ensemble.
This is repeated until a new reactive trajectory is harvested~\cite{Li_Ma2016TPS}. Each reactive trajectory is $2.5$ ps long, with the time step of $1$ fs.  We then collect the configuration (conformation and velocity) at every step along each trajectory. All together, we have $7.5\times 10^{9}$ conformations.

\paragraph{Constructing Time-Evolving Dynamic Probability Surface.} 
We align the trajectories by the time of the occurrence of the transition, with the time $t$ at transition set to $t=0$.  Conformations at the transition state have the appropriate values of the one-dimensional reaction coordinate as described in~\cite{wu2021mechanism}. After alignment, we examine the  time-interval of transition from $-2.5$ ps to $+2.5$ ps. 
We construct the time-evolving dynamic probability surface 
$\{ p(\bx,\, t) | (\bx,\, t) \in \mathbb{M}_t\}$ 
using the $7.5\times 10^{9}$ aligned and time-stamped conformations. 
Based on the analysis of reaction coordinates using the energy flow theory~\cite{Li_Ma2016_Reaction_Mechanism}, we select the top-ranked $5$ dihedral angles ($\phi$, $\psi$, $\theta_1$, $\alpha$, $\beta$) from the original $60$ spatial dimensions as the coordinates of $\mathbb{M}_s$. 
Along with  time $t$, we have a $6$-dimensional probability surface $\{ p_t(\bx,\, t) | \bx \in \mathbb{M}_t\}$. 
Each angle coordinate of ($\phi$, $\psi$, $\theta_1$, $\alpha$, $\beta$) 
in units of radians
is divided into 15 bins, and the time interval is divided into 500 bins, each of $10$ fs. 
This discretization leads to to $15^5\times 500= 379,687,500$ 6-dimensional hypercubes, where time is one of the dimension.

\paragraph{Computing Topological Structure of the Dynamic Probability Surface.} Persistent homology is computed using a $20$-core Xeon E5-2670CPU of 2.5 GHz, with a cache size of 20 MB and memory of 128 GB Ram. The computation time for finding the prominent peaks and ridges connecting them is about $10$ min.

\subsection{High Probability Reactive Region in Space-Time from Topology of Dynamic Probability Surface} \label{Surf_topo}

\begin{figure}[!htbp]
    \includegraphics[width=0.98\linewidth]{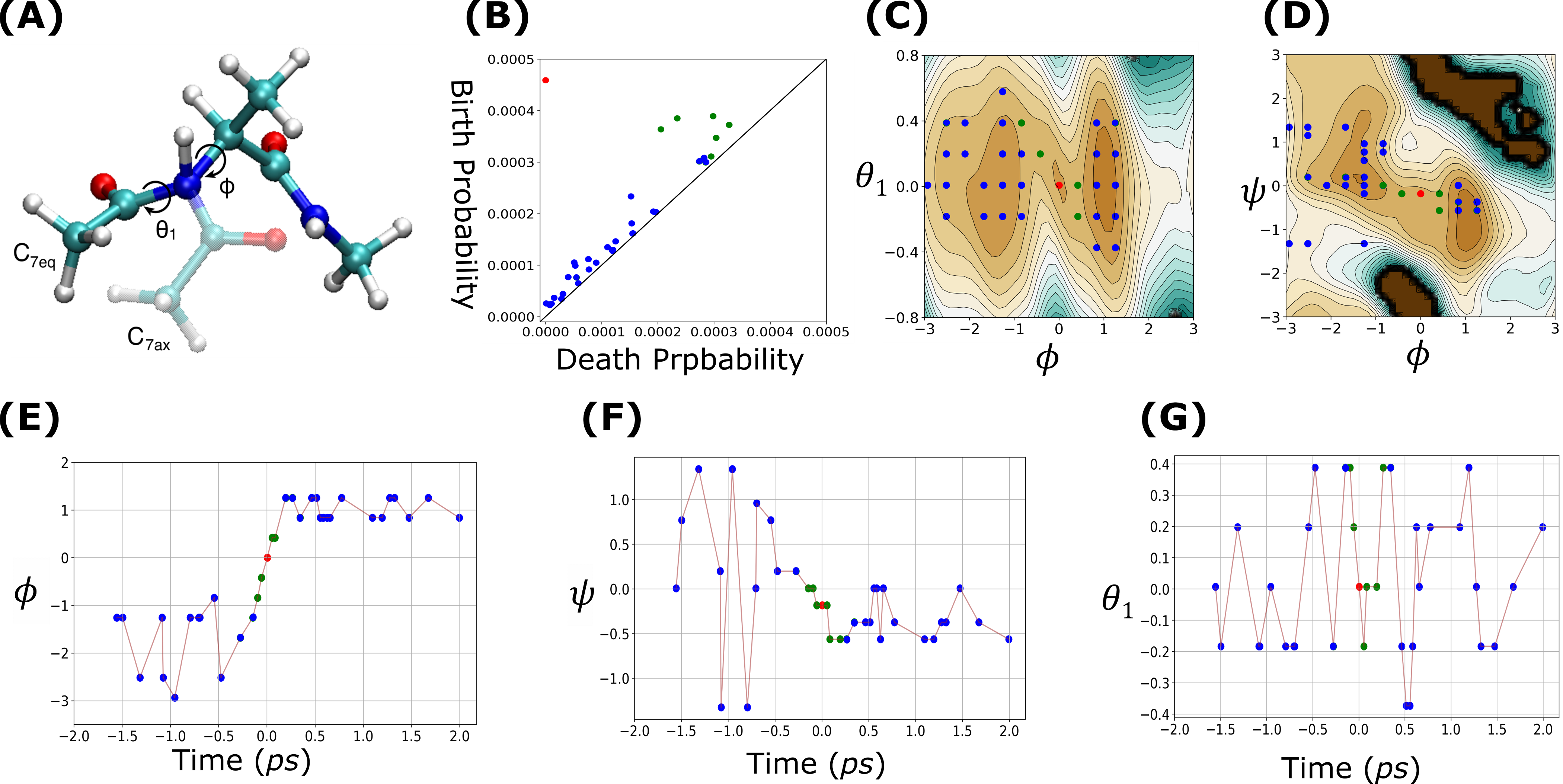}
    
    \caption{\sf The dynamic probability landscape and its topological structure.
    (\textbf{A}) 3-d conformation of the alanine dipeptide before and after isomerization. 
    (\textbf{B}) The persistence diagram of probability peaks over the (time, $\phi, \psi, \theta_1, \alpha, \beta$)-space, where the birth and death probability of each peak is shown.
    (\textbf{C}) The 6-d landscape projected onto the $\phi$--$\theta_1$ plane. Here colored dots are the locations of probability peaks occurring at different time. 
    The contour plot in the background depict the sea level of time-aggregated probability projected onto this 2-d plane, where brown and cyan color indicates high and low probability, respectively.
    (\textbf{D}) The landscape projected onto the $\phi$--$\psi$ plane. 
    (\textbf{E}) The $\phi$ coordinate of the probability peaks as time proceeds.  $\phi$ fluctuates in the reactant basin before transition occurs. During the transition period ($t=0$), $\phi$ increases and subsequently  reaches to the value of the product basin. 
    (\textbf{F}) The $\psi$ coordinate of the probability peaks as time proceeds. There is significant fluctuation in $\psi$ before transition.
    (\textbf{G}) The $\theta_1$ coordinate of the probability peaks as time proceeds. $\theta_1$ fluctuates throughout the whole time.
    As there are only a finite number of trajectories, we coarse grained each coordinate into 15 bins.  As probability peaks in 6-D
    space are shown on 2D-planes, separate probability peaks in space or time may appear at the same coarse-grained locations in these 2D angle plots. The location of each peak, and its birth probability is shown in Supporting information table S1.
    }
    \label{fig:topo_TimeAnd3}
\end{figure}

\paragraph{High probability reactive region dominates in the configuration-time space during the transition.}
In a previous study, we showed that without time separation, the transition state conformations among the aggregation of 
$7.5\times 10^{9}$ conformations during the  period of $-2.5$ to $+2.5$ ps are
concentrated in a small reactive region of $\phi \times  \theta_1 = [-0.2,\, +0.2] \times [-0.1,\, +0.1]$ (see Fig~\ref{fig:topo_TimeAnd3}C for the $\phi$ and $\theta_1$ angles).  
These reactive conformations pass the rigorous committer test and are at the transition state, and  they form the most prominent  peak with the largest probability mass outside the reactant and product basins~\cite{Manuchehrfar_Activated}.

With time as the extra dimension, we now examine the detailed time sequences of the probability surface and determine how transition state conformations are distributed during this period of $5$ ps.  This is captured by the 6-d space-time probability surface and its overall topological structure 
 is summarized in the persistent diagram~(Fig~\ref{fig:topo_TimeAnd3}B), with  the projections of the surface on the $\phi$--$\theta_1$ and $\phi$--$\psi$ planes shown in Fig~\ref{fig:topo_TimeAnd3}C-\ref{fig:topo_TimeAnd3}D. 

One prominent probability peak (Fig~\ref{fig:topo_TimeAnd3}B, red dot) located  in the region of 
$(\phi,\,\theta_1) = [-0.2,\, +0.2]\times[-0.1,\, 0.1]$ stands out, which occurs during the short time interval of $t \in [-5,\, +5]$ fs.
This is the reactive region where most probability mass of  the transition state conformations accumulates. 
It forms the dominating topological structure with the largest persistence in the whole configuration-time space.  
The  probability peak of time-aggregated conformations from  $-2.5$ to $+2.5$ ps reported in ref~\cite{Manuchehrfar_Activated}
largely arises from this short-durationed reactive probability peak. 
That is, the dominant peak of ref~\cite{Manuchehrfar_Activated} comes from the dominant peak occurring during  $t = [-5,\, +5]$ fs reported here.

There are 6 additional meta peaks at the next level of probability height (green dots) with much smaller persistence. Most of these occur near the transition state  within $\pm 0.2$ ps from $t=0$.  The remaining 54 peaks are near either the reactant basins or the product basins, reflecting minor fluctuations within these stable regions.

\paragraph{Probability peaks in the reactive region at the transition time.}
The locations of the probability peaks in
 $\phi$, $\psi$, and $\theta_1$ 
 against time $t$ are shown in
 Fig.~\ref{fig:topo_TimeAnd3}E--\ref{fig:topo_TimeAnd3}G. 
In the $\phi$ angle, minor probability peaks fluctuate around the reactant basin ($\phi \approx -2.0$) rad 
before reaching the transition state (Fig~\ref{fig:topo_TimeAnd3}E).
At $<-0.5$ ps prior to the transition state, $\phi$ increases rapidly to the value of the product basin
($\phi\approx1.0$) rad, and  fluctuates modestly afterwards.  At $t=0$, the highest probability peak (red dot) occurs at $\phi=0$. 
The $\psi$ angle fluctuates drastically around the reactant basin ($\psi \approx 0$) prior to the transition state (Fig~\ref{fig:topo_TimeAnd3}F). 
Near the transition  $t=0$, $\psi$  decreases gradually to the value of product basin
($\psi\approx 0.5$) rad and become more stabilized after the isomerization. 
In the reaction coordinate $\theta_1$, the probability peaks fluctuates significantly  in a consistent manner throughout the entire $5$ ps, exhibiting an overall oscillating  behavior (Fig.~\ref{fig:topo_TimeAnd3}G).

The probability peaks are all small before the transition (blue), reflecting the fact that the molecular conformations prior  to isomerization are diverse.  
At the transition time $t=0$,  a probability peak is located in the small region of $\phi \times \theta_1 \times \psi = [-0.2,\, +0.2]\times[-0.1,\, +0.1]\times[-0.2,\, +0.2]$ (Fig.~\ref{fig:topo_TimeAnd3}E-\ref{fig:topo_TimeAnd3}G, red). 
This reflects the fact that most conformations on route to isomerization pass through a small reactive region in the configuration space.  After the transition, probability peaks  again become small (blue), reflecting the diverse conformations near the product basin.

Overall, these results show that the transition state at $t=0$ has the highest probability peak, which is preceded  and followed by smaller meta peaks (two before and seven after $t=0$),  all within $\approx \pm0.4$ ps of the transition time.  
At the reactant and product basins,  molecular  conformations are diverse, with a number of small probability peaks.
As the probability peak increases then decreases in height, there is 
 consistent fluctuation in the $\theta_1$ angle , while   
moderate fluctuations occur  before transition in $\phi$ and in $\psi$.

\paragraph{Relation between free energy surface and the dynamic probability surface.}

\begin{figure}[!htbp]
    \includegraphics[width=0.98\linewidth]{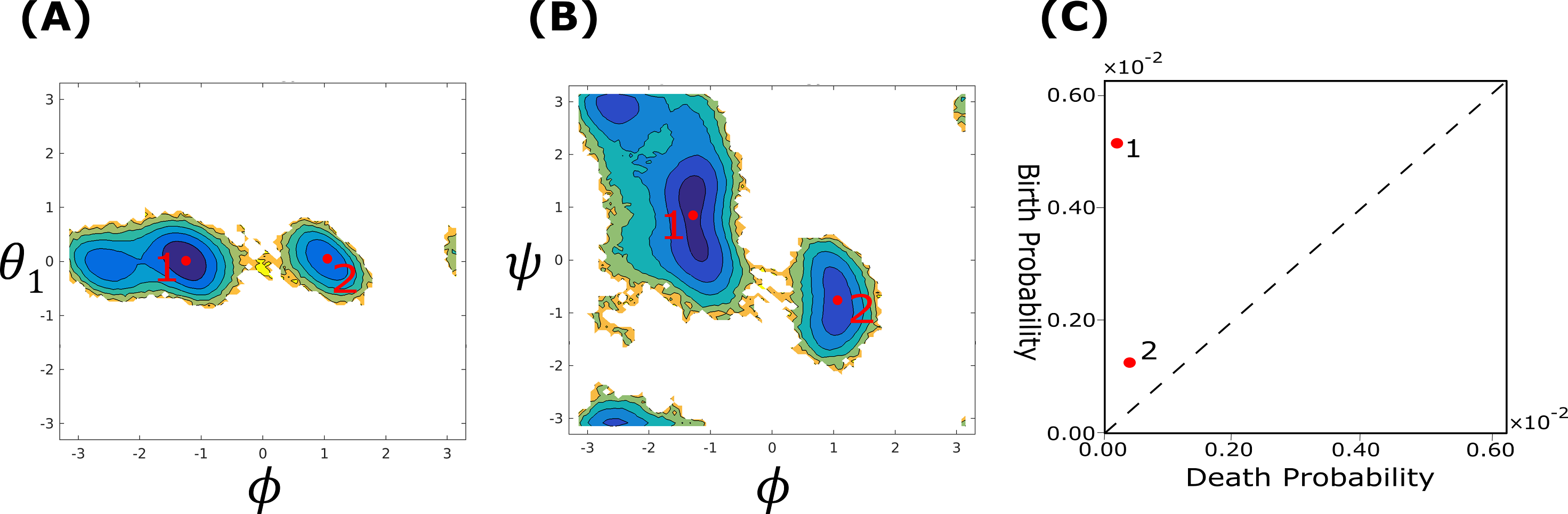}
    
    \caption{Free energy surface approximated from a long MD trajectory of $15$ ns plotted on {\bf (A)} the $\phi$--$\theta_1$ plane and the {\bf (B)} $\phi$--$\psi$ plane. Its persistent diagram {\bf (C)} shows that there are two prominent minima on the free energy surface, or peaks on the probability landscape, which are associated with the reactant basin (labeled 1) and the product basin (labeled 2).}
    
    \label{fig:ActualPotential}
\end{figure}

The potential energy surface of the alanine dipeptide isomerization in vacuum is as previously described in Ref.~\cite{Bolhuis2000}.  There are two prominent minima on the potential energy surface, associated with the reactant basin and the product basin. Their locations are identical to the locations of the reactant and the product basin on the dynamic probability surface reported in~\cite{Manuchehrfar_Activated}. We have also determined the location of minima on the free energy surface, which are derived 
from a longer MD trajectory of $\approx 15$ ns. The $3.0 \times 10^7$ conformations taken from each $0.5$ fs intervals of the trajectory are harvested, from which the free energy surface is approximated~(Fig.~\ref{fig:ActualPotential}A and \ref{fig:ActualPotential}B).
The topological structure of the free energy surface is summarized in the persistent diagram of Fig.~\ref{fig:ActualPotential}C. There are two prominent minima on the free energy surface, or equivalently, two high probability peaks on the probability surface, when examined over the 3-dimensional $\phi$--$\psi$--$\theta_1$  space 
(red dots on Fig.~\ref{fig:ActualPotential}A and \ref{fig:ActualPotential}B). One is associated with the reactant basin (($\phi$, $\psi$, $\theta_1$) $=$ $(1.25,\,-0.84,\,0.01)$ (labeled 1),
and the other with the product basin $(-1.68,\,0.42,\,-0.18)$ (labeled 2).  These locations 
are identical to the locations  of the reactant and the product basins on the dynamics probability surface of reactive trajectories as reported in~\cite{Manuchehrfar_Activated}. 
     Furthermore, 
 there exists a probability peak located at the active region of ($\phi$, $\psi$, $\theta_1$) $=$ $(0.00,\,-0.42,\,0.00)$ on the
dynamic probability surfaces, regardless whether it is time-separated as discussed earlier or 
over the whole $2.5$ ps period~\cite{Manuchehrfar_Activated}.
However, there is no corresponding minimum on the free energy surface at this location.

\subsection{Reactive Vortex Regions of 
High Probability Exhibit Strong
Non-Diffusive  Rotational Flux}

\begin{figure}[!htbp]
    \includegraphics[width=0.98\linewidth]{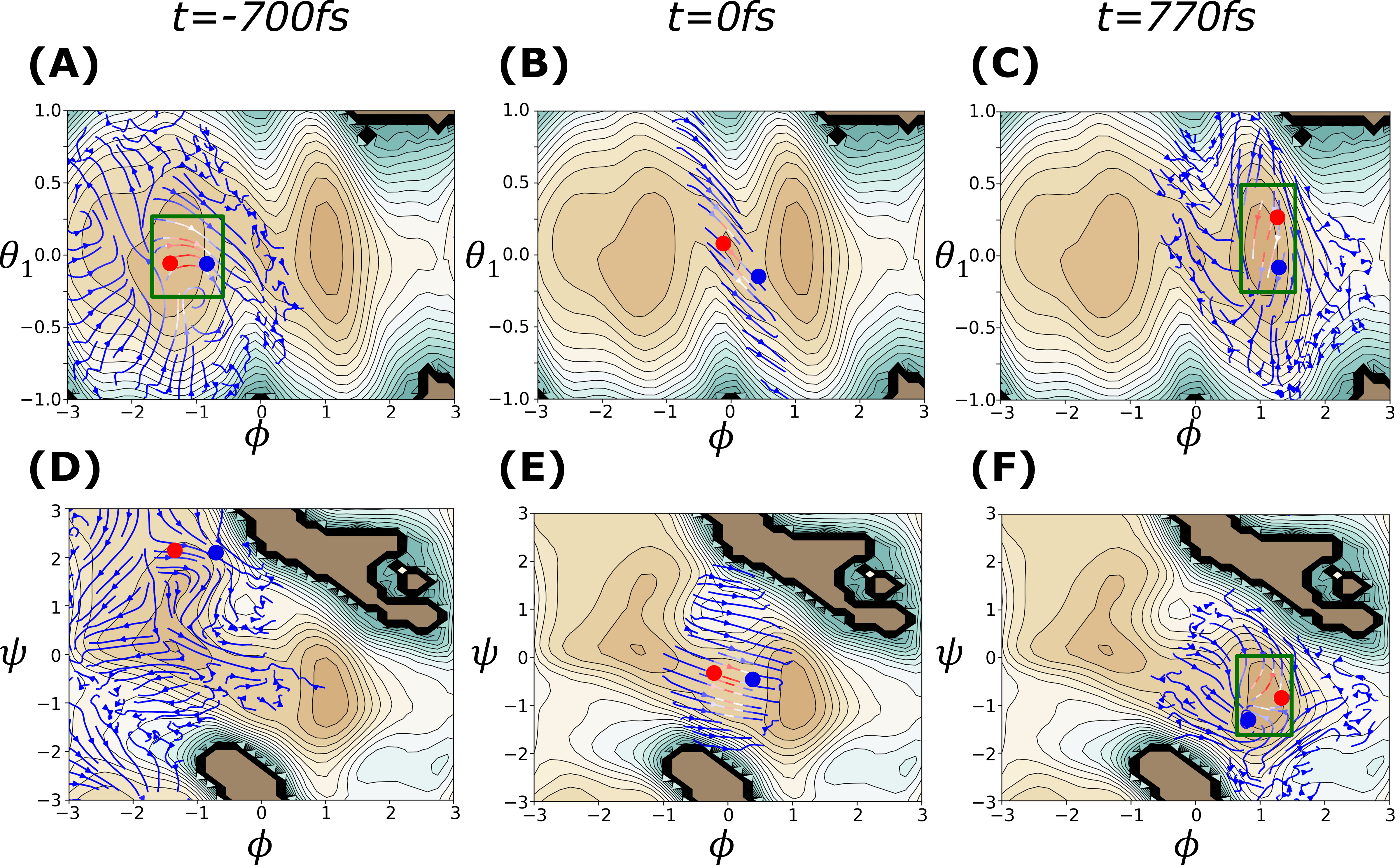}
    
    \caption{\sf Dynamic fluxes projected on the $\phi$--$\theta_1$ and the $\phi$--$\psi$ planes at three different times of before ($t=-700$ fs), at ($t=0$ fs), and after ($t=+770$ fs) the transition. The strongest portions of the flux lines are in red. Red dots are locations of probability peaks at the current time, and blue dots are the 
    location of peaks after $20$ fs.}
    \label{fig:flux_TimeAnd3}
\end{figure}

\paragraph{Flux and projection to the $\phi$--$\theta_1$ and $\phi$--$\psi$ planes.} 
We study  dynamic fluxes of molecular movement, which is calculated using Eqn~(\ref{eqn:flux-disc}).
We first study the projection of the flux lines 
to the $\phi$--$\theta_1$  and $\phi$--$\psi$  planes (Fig~\ref{fig:flux_TimeAnd3} and Supplementary movies~2-3), and examine how they are related to topological changes in the probability peaks 
on the 6-d space of (time$, \, \phi,\, \psi, \, \theta_1, \, \alpha,\, \beta$).
For illustration, we take 3 time points before ($t= -700$ fs), at  ($t= 0$ fs), and after ($t= +770$ fs) the transition. 

At $t=-700$ fs, flux is  present in the cubic region of  $\theta_1 \in [-1.0, \, +1.0]$, \, $\phi \in [-3.0,  \, 0.5]$, and $\psi \in [-3.0, \, +3.0 \, ]$. 
Upon projection onto the $\phi$--$\theta_1$ plane, strong and uneven fluxes are located  in a smaller rectangle of $\theta_1 \in [-0.2, \, +0.2]$ and $\phi \in [-1.5, \, -0.5]$ (green rectangular region in 
Fig.~\ref{fig:flux_TimeAnd3}A  and  red flux lines in Supplementary movie 2).
This is the same location where probability peak  at   $t=-700$ fs (Fig~\ref{fig:topo_TimeAnd3}D and in Fig~\ref{fig:topo_TimeAnd3}A). 
 When projected to the $\phi$--$\psi$ plane, the flux is weak and even-valued (Fig.~\ref{fig:flux_TimeAnd3}D and Supplementary movie 3).

At the transition time $t=0$, flux lines are the strongest on both the
$\phi-\theta_1$ and the $\phi-\psi$ plane. They are in the direction of increasing $\phi$ and decreasing $\theta_1$, and slightly decreasing $\psi$ (Fig.~\ref{fig:flux_TimeAnd3}B, \ref{fig:flux_TimeAnd3}E, and Supplementary movies~2-3). 
This is the direction pointing from the reactant basin to the product basin.
The probability peak at $t=0$ is located at the center of the flux lines (red dots in Fig.~\ref{fig:flux_TimeAnd3}B and~\ref{fig:flux_TimeAnd3}E).

At $t=770$ fs after the transition, dynamic flux is found in the cubic region of  $\phi \in [-0.5, \, 2.0]$, $\theta_1 \in [-0.75, \, 0.75]$, and $\psi \in [-3.0, \, 1.5]$~(Fig~\ref{fig:flux_TimeAnd3}C and ~\ref{fig:flux_TimeAnd3}F). 
When projected onto the $\phi$--$\theta_1$ plane, the flux is uneven and is the strongest 
around the rectangle of $\phi \in [0.8, \, 1.2]$ and $\theta_1 \in [-0.2, \, 0.5]$
(green rectangle, Fig~\ref{fig:flux_TimeAnd3}C).
It is also uneven in the $\phi$--$\psi$ plane and  is the strongest around the rectangle of $\phi \in [0.8, \, 1.2]$ and $\psi \in [-1.5, \, 0.0]$~(green rectangle, Fig~\ref{fig:flux_TimeAnd3}F).

Overall, these results show that the directional fluxes of molecular movement emerge during the transition period.
Fluxes are concentrated in the high probability reactive region in the configuration-time  space
and drive the probability peak of molecular configurations to future locations (red to blue dots, Fig.~\ref{fig:flux_TimeAnd3}A-~\ref{fig:flux_TimeAnd3}F).
At the  transition time, they are the strongest and 
 are in the general direction of moving molecules towards the product basin.

\paragraph{The reactive vortex region has strong rotational flux during transition.} 
We further study the rotational flux of molecular movements  during the transition.
Its  projections onto the  $\phi$--$\theta_1$ and the $\phi$--$\psi$ planes are
$({\partial J_{\Delta\Omega,\theta_1}(t)}/{\partial \phi} - {\partial J_{\Delta\Omega,\phi}(t)}/{\partial \theta_1})$
and
$({\partial J_{\Delta\Omega,\psi}(t)}/{\partial \phi} - {\partial J_{\Delta\Omega,\phi}(t)}/{\partial \psi})$, respectively~(Eqn~(\ref{eqn:rot})).

\begin{figure}[!htbp]
    \includegraphics[width=0.98\linewidth]{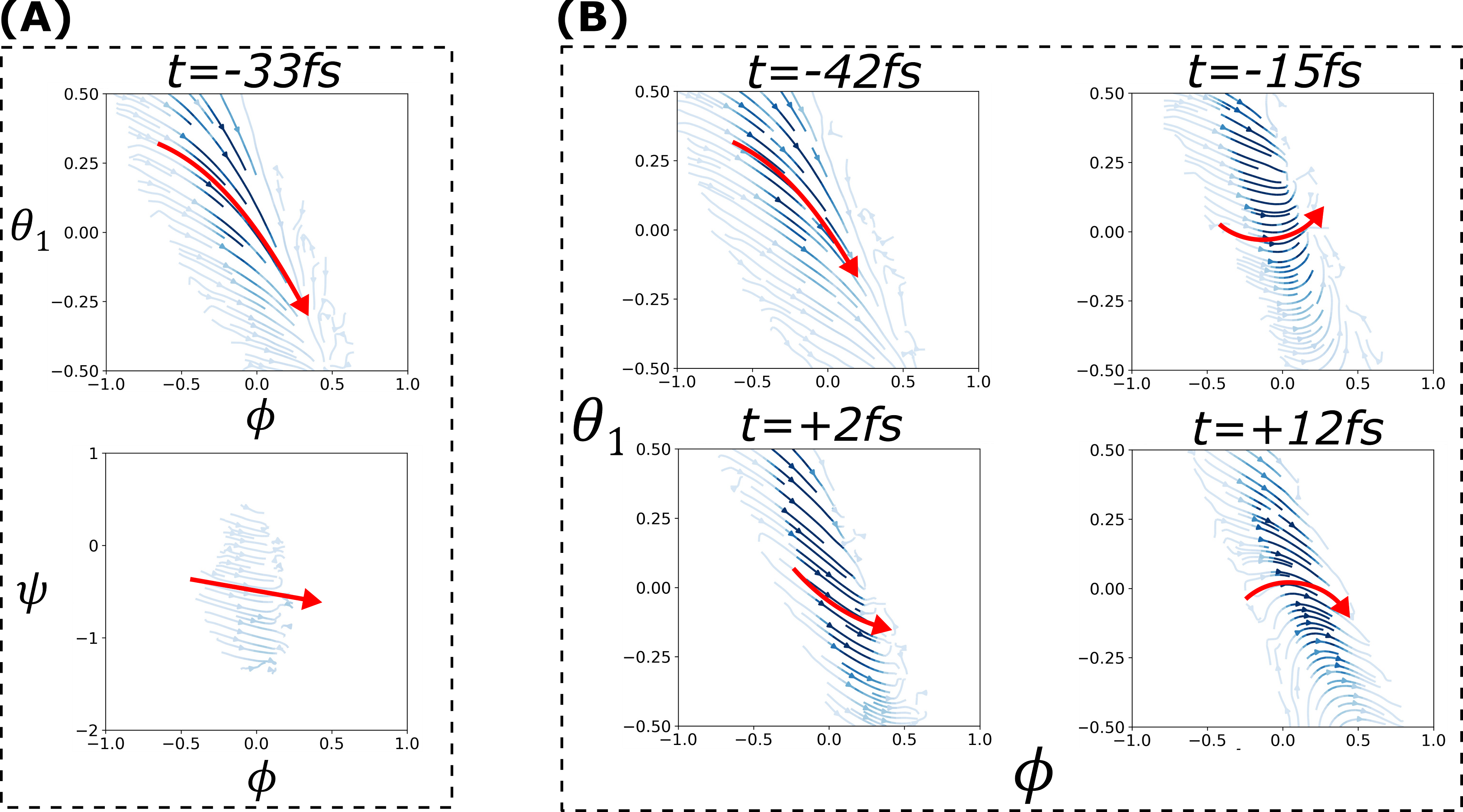}
    \caption{\sf The flux and its rotation during the transition at $t=-33$ fs~(A), $t= -42$ fs, $-15$ fs, $+2$ fs, and $+12$ fs~(B). 
    The flux lines are shown as blue lines, with the flux rotation coded by the color intensity, where darker blue represents stronger rotation. 
    (A) There is strong flux rotation in the plane of the two reaction coordinates of $\theta_1$ and $\phi$  (darker blue) at $t=-33$ fs (top), while flux rotation is negligible in the plane of $\psi$ and $\phi$ (bottom).
    (B) Strong flux rotation presents at $t=-42$ fs, $-15$ fs, $+2$ fs, and $t= +12$ fs as the flux lines changes direction. In contrast, flux rotation remains negligible in the $\phi$--$\psi$ plane (supplementary movie 4), with the flux maintaining the same direction as in (A).
    Arrows in red represent the overall directions of the flux lines.
    }
    \label{fig:HighResFlux_TimeAnd3}
\end{figure}

We focus on the high probability reactive region and  examine  the rotational flux  during the time interval of $-50$ fs and $+50$ fs in the reactive cubical region where 
most  probability mass is located~(Fig~\ref{fig:HighResFlux_TimeAnd3}).
We divide the interval in each dimension of the cube $\phi \times \theta_1 \times \psi \in [-1.0,\, 1.0]
 \times[-0.5,\, 0.5]\times[-2.0,\, 1.0]$ containing the reactive region into 250 bins and examine the flux and rotational flux in the $250^3=15,625,000$ cubes. 

There are strong rotational fluxes in the $\phi$--$\theta_1$  plane, the two most important reaction coordinates~(Fig.~\ref{fig:HighResFlux_TimeAnd3}A, top). 
The flux exhibits significant changes in the direction of $\theta_1$ as time proceeds while maintaining the same direction of increasing $\phi$ (curved arrows, Fig.~\ref{fig:HighResFlux_TimeAnd3}B, top, and supplementary movie 4). 
In contrast, 
 flux lines in the $\phi$--$\psi$ plane moves along a fixed direction of increasing $\phi$ and decreasing $\psi$. The rotational flux in this plane is  negligible, even though $\psi$ is the important coordinate that  defines geometrically the reactant and product basin along with $\phi$~(Fig.~\ref{fig:HighResFlux_TimeAnd3}A, bottom). 
Overall, these results show there are strong vortexes in  the reactive region.

\begin{figure}[!htbp]
\centering
    \includegraphics[width=0.38\linewidth]{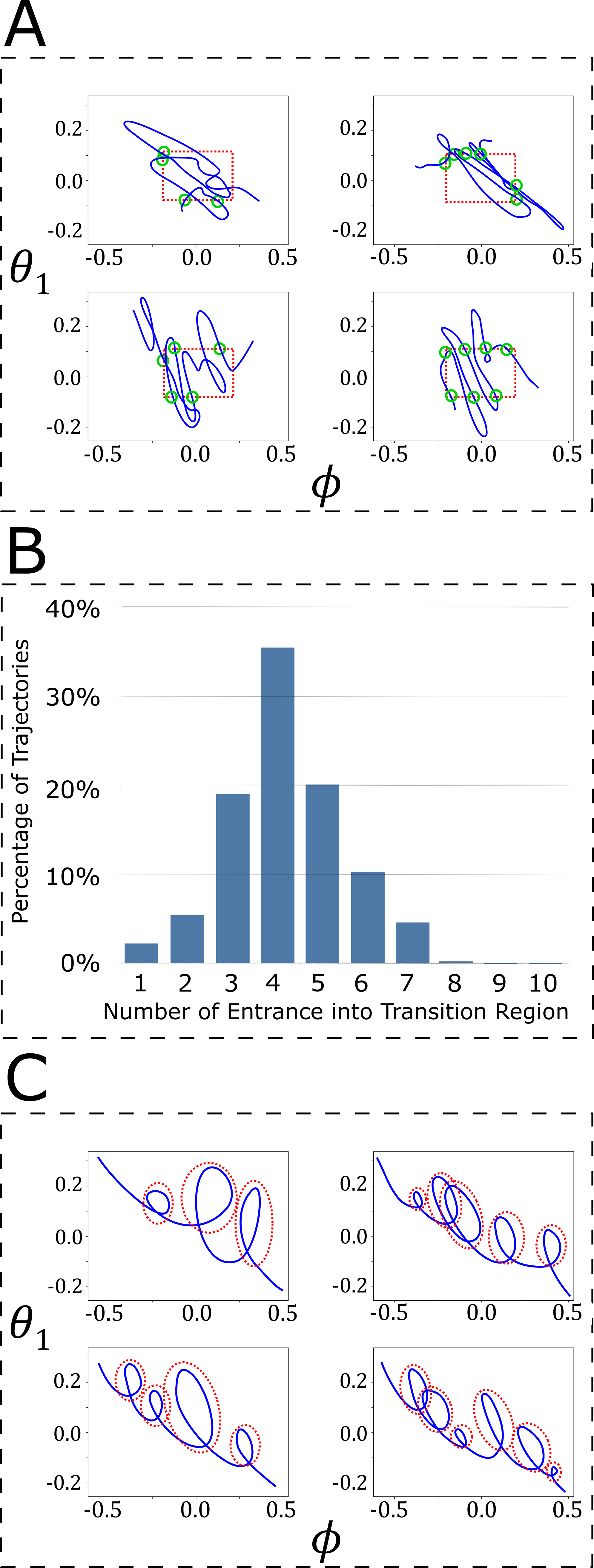}
    \caption{\sf Rotating trajectories in the reactive vortex region. 
    \textbf{(A)} Examples of trajectories which enter the reactive region one, two, four, and six times. Here the reactive region of transition state region is indicated by the dashed red rectangle of $(\phi, \theta_1)=[-0.2,\, +0.2]\times[-0.1,\, +0.1]$ as discussed in Fig.~\ref{fig:topo_TimeAnd3} 
    and in~\cite{Manuchehrfar_Activated}.
    Each entrance/reentrance point of a trajectory is highlighted by a green circle.
    \textbf{(B)} Distribution of the number of entrance of rotating trajectories exhibit during the transition. Majority of trajectories enter into the reactive vortex region between three and five times, with a small proportion  rotate more than six  or less than  three times.
    \textbf{(C)} Additional examples of trajectories circulate three, four, five, and six times around the reactive vortex region. Here dashed red circles highlight circles in each trajectory.}
    \label{fig:rotation}
\end{figure}

\paragraph{Non-diffusive rotational trajectories in reactive vortex region dominate the transition process.} Our above analysis of flux over time are at $10$ fs resolution. As  trajectories of molecular movement pass through the transition state rapidly, we now examine the behavior of trajectories during transition at finer resolution of $1$ fs. To gain further insight into the reactive vortex region, we study the behavior of trajectories of molecular movement and measure the number of times that each trajectory rotates during the short transition time interval of $[-100,\, +100]$ fs. This is calculated by counting the number of times a trajectory re-enters the transition state region. 
Here we regard the small rectangle   of $(\phi, \theta_1)=[-0.2,\, 0.2]\times[-0.1,\, 0.1]$ as the reactive region, which is where the red dots in Fig.~5C and 5D are located. Results in ref~\cite{Manuchehrfar_Activated} showed 
that conformations of transition state ensemble 
are indeed located in this region,  as these conformations pass the rigorous committor test (dashed red rectangle, Fig~\ref{fig:rotation}A, see also discussion related to Fig.~\ref{fig:topo_TimeAnd3}).
Fig.~\ref{fig:rotation}A shows example trajectories  with different number of entrance and re-entrance to the reactive region. 

The distribution of the number of times that trajectories enter the transition region is shown in Fig.~\ref{fig:rotation}B  for a sample of 100,000 trajectories.
The majority of them re-enter the transition region 3--6 times, with those re-enters 4 times occurring most frequently  ($35.2\%$).  
 Trajectories entering the transition region 5, 3, and 6 times represent
$20.0\%$, $18.6\%$ and $10.7\%$ of the sampled trajectories, respectively.
Fig.~\ref{fig:rotation}C show additional examples of
trajectories rotating inside the transition region moving in well-formed circles (3, 4, 5, and 6  times, respectively).

These results show that most trajectories in the reactive vortex region rotate multiple times, exhibiting strong  non-diffusive rotational dynamics.  There is a broad distribution in the number of re-entrance into the transition-state region, with the majority of them experiencing 3--6 rounds of rotations.  

\paragraph{Rotational fluxes are important  for  barrier crossing.}
The flux lines  in Fig~\ref{fig:HighResFlux_TimeAnd3}B (e.g., $t=-33,\, -42$ and $2$ fs)  show that molecules generally move along in the direction that coincides with the isosurfaces of ensembles of conformations with the same committor value as described in ref~\cite{wu2022rigorous}.
The rotational flux carries the molecules in the  direction orthogonal to the direction of the isocommittor, indicating barrier crossing. 
As representatives of most trajectories, the examples in Fig~\ref{fig:rotation}A  show that molecules
move rapidly in the general direction of isocommittor surfaces, but slowly in the orthogonal direction towards other isocommittor surfaces.  The combination of these movements result in an overall spiral-like trajectories with elongated ellipse(s) drawn-out in the projection of $\phi$--$\theta_1$ reaction coordinates.
During barrier crossing, the motion in $\phi$ is assisted by $\theta_1$,  which transfers the potential energy it received from the thermal bath to $\phi$  directly via kinetic energy as discussed in ref~\cite{wu2022rigorous}. This leads to tight cooperative movements between $\theta_1$ and $\phi$, which is manifested as rotational flux.  

\section {Discussion}
The transitions state theory has been the corner stone for understanding  activated processes ranging from isomerization of simple organic molecules to complex  protein conformational changes.  Central to this theory 
are the transition state ensemble, namely, molecular conformations at the barrier top occupying the 1-degree saddle point of the free energy surface. 

Dynamics of the transition state ensemble along the reaction coordinates is an important component of reaction rate theories. The default assumption for complex systems was based on Kramers’ physical intuition rather than systematic examination of the transition dynamics in realistic systems. As a result, the dynamics of transition of naturally occurring activated processes in complex molecules are largely unexplored.

In this study, we quantified the detailed topological structures of the dynamic probability surface of an activated process over the time-configuration space.
We use  the alanine dipeptide isomerization in vacuum as our model system.  The dynamic probability surface is constructed by harvesting naturally occurring trajectories of molecular movements connecting the reactant and the product basins.
Unlike small molecules that require an external energy source, alanine dipeptide is the smallest complex system with an internal heat-bath composed of the large number of non-reaction coordinates. This heat-bath provides the necessary energy flow to facilitate the barrier-crossing process, an important property shared by proteins but absent in small molecules.

Our results are based on rigorous analysis of the topological structures of the high-dimensional dynamic surfaces using persistent homology.  In addition,  we  introduce a new method for quantifying high-dimensional flux rotations. 
These techniques allowed us to uncover a number of important insights.

First, the transition state ensemble of conformations are located in a reactive region in the configuration-time space and form the dominant probability peak.  This finding extends earlier results of ref~\cite{Manuchehrfar_Activated} and shows that after further separation of transition-state conformations along the time-axis, a single prominent probability peak occurring
during the short interval $t = [-5,\, +5]$ fs
dominate throughout the transition barrier-crossing process.  That is,  a strong reactive region with the highest probability peak exists in configuration-time, where transition state conformations as verified by the rigorous committor test accumulate are located.  This region 
of short time duration dominates the whole transition process.

Second, there are strong directional fluxes in the high-probability reactive region. Molecules in this active region are not in equilibrium and are not diffusion-controlled.    The fluxes adjust directions and become uniformly aligned at the transition time when they are the strongest, with the probability peak located at the center of flux lines. These fluxes occur primarily in the subspace of the reaction coordinates, and  carry the molecular conformations forward.

Third, the reactive region is characterized by strong vortexes. 
There are strong rotational fluxes at the transition state, which occur in the subspace of the two most important reaction coordinates,  but not in the subspace of the most important geometric coordinates.  Most trajectories on route to the product basin rotate and enter the reactive vortex region multiple times.  
These reactive trajectories move along rapidly in the direction of the surfaces of isocommittors, but slowly in the orthogonal direction  to scale the barrier to the next isocommittor surface,
 drawing out spiral-like curves encircling ellipses elongated in the direction of the isocommittor surfaces. 
The tight cooperative movements
between reaction coordinate $\theta_1$ and $\phi$  are due to the transfer of potential energy $\theta_1$ received from the thermal bath to $\phi$. 
The dynamic movements along the isocommittor surface and  in the orthogonal direction of barrier-crossing are manifested as rotational fluxes in the plane of the reaction coordinates.

Overall, our findings offers a first glimpse 
into the reactive vortex region that characterizes the non-diffusive dynamics  of  barrier-crossing  of a naturally occurring activation process.
By separating conformations along the time axis, we uncovered rich topological structures in the dynamic probability surface.
Such details are not possible when examining the free-energy surface and its 1-saddle point, where the dynamic aspects of the process are obscured.

The discovery of the reactive vortex region highlights the importance of analyzing the topological structures  of the dynamics of the transition region in naturally occurring  activated processes.  With alanine dipeptide being the first system where non-diffusive behavior is established, it will be fruitful to study reactive dynamics of  other naturally occurring activated processes of complex molecules.  
The results can serve as the  foundation towards developing a theoretical model of transition dynamics describing activated process occurring in nature.

While our study does not directly provide physical quantities such as rate constants that correspond to experimental measurements, it is possible in principle  to analyze how fluxes crosses  dividing separatrix surface and to estimate the reaction rate as described in~\cite{rosenberg1980isomerization,bose2017non,jang1992comment,zhao1993comment,nagahata2021phase}, provided one can precisely define the
separatrix surface and can accurate sample and quantify the fluxes.

\section*{Acknowledgement}
We thank Drs.\ Hubert Wagner and Herbert Edelsbrunner for their generous help in extending the cubic complex algorithm. We also thank Dr.~Wei Tian for his help. 
This work is supported by grants NIH R35
GM127084 (to JL), NIH R01 GM086536 (to AM), and NSF CHE-1665104 (to AM).

\section*{Conflict of Interest Statement} 

There are no conflict of interests.

\bibliography{Reference}

\newpage

\section{Supporting Information}

\subsection{Topology of time-evolving probability surface on $\alpha$--$\beta$  planes.}
\begin{figure}[!htbp]
    \includegraphics[width=0.98\linewidth]{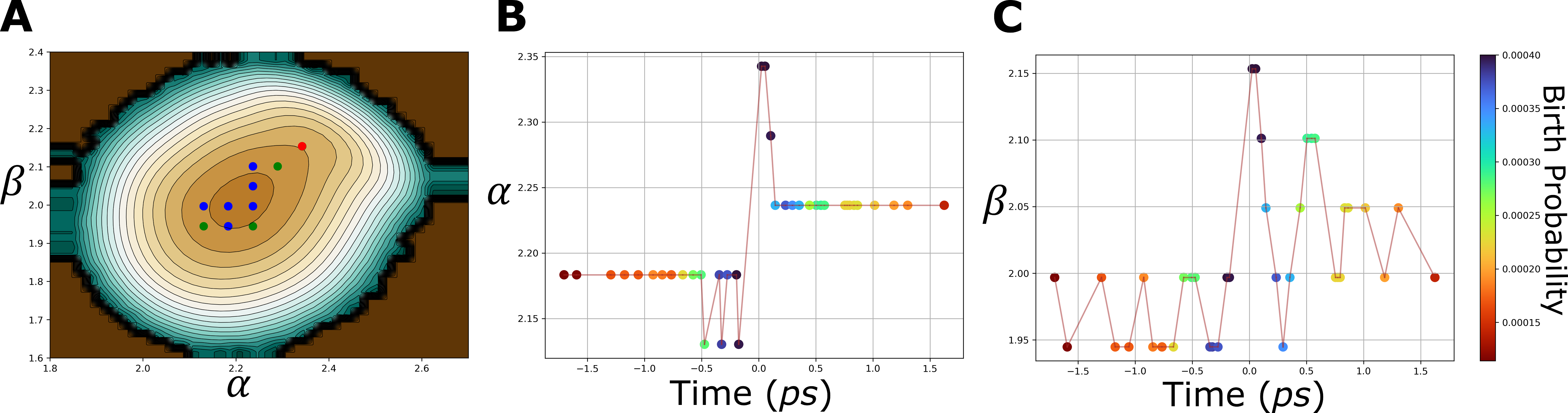}
    
    \caption{ Topological structure of the dynamic probability landscape.
    \textbf{A} The dynamic probability landscape projected onto the $\alpha$--$\beta$ plane. dots are the location of peaks at different time. 
    \textbf{B} The $\alpha$ coordinate of peaks as time proceeds.  $\alpha$ is small before transition, reflecting the location of reactant basin.  The value of $\alpha$ fluctuates during the transition period and increases to its maximum at  transition time of $t=0$. After transition, $\alpha$ decreases into the value of the product basin, which is larger than that of the reactant basin.
    \textbf{C} The $\beta$ coordinate of peaks as time proceeds. It shows that before transition time ($t=0fs$), 
    $\beta$ fluctuated with a small amplitude right before the transition, and increases to its maximum value at transition ($t=0$). After transition, $\beta$ fluctuates at a  small amplitude and stabilizes at a larger value corresponding to the product basin.}
    
    \label{fig:topo_Time_alpha_beta}
\end{figure}

We then examine the topological structures of the $6$-d time-evolving dynamic probability surface on $(\alpha, \beta)$ plane.
$\alpha$ and $\beta$ are not important reaction coordinates as they rank the $4$-th and $5$-th in their importance, respectively. 
Fig~\ref{fig:topo_Time_alpha_beta}A depicts this  dynamic probability surface on the $\alpha$--$\beta$ plane and the location of the probability peaks on this plane.
The birth and death probability of each of these peaks were shown in the persistent diagram (Fig.~5B).
The transition state is the most prominent peak with the longest persistent. The location of the transition state peak on $\alpha$--$\beta$ plane is shown as a red point in Fig~\ref{fig:topo_Time_alpha_beta}A. It is followed by nine meta-stable peaks, in which their location is shown in green on Fig~\ref{fig:topo_Time_alpha_beta}A. There are several minor peaks in the persistent homology analysis, which are shown as blue point in Fig~\ref{fig:topo_Time_alpha_beta}A. 

We then  ordered these prominent peaks by their occurrence time, and plot their $\alpha$ and $\beta$ locations over time in separate plots (Fig.~\ref{fig:topo_Time_alpha_beta}B-\ref{fig:topo_Time_alpha_beta}C). 
As shown in Fig.~\ref{fig:topo_Time_alpha_beta}C, probability peaks in $\alpha$ angle is constant before the transition state at  $\approx 2.18$, and reaches a different constant angle at $\approx 2.34$ after the transition. In the transition period ($-0.5$ -- $+0.2 ps$), $\alpha$-angle fluctuates, during which it reaches the maximum angle of $\approx 2.34$.
%
%
%
The probability peaks vary in $\beta$ within a small interval of 1.95--2.00 before the transition, and stabilize between 2.00--2.05 after the transition, while reaching the maximum $\beta$ angle of 2.15 during the transition.

Overall, $\alpha$ and $\beta$ do not experience significant changes before and after the transition, as differences in $\alpha$ and $\beta$ between the reactant and product basins are quite small. During the transition period, changes in both $\alpha$ and $\beta$ are larger.

\subsection{Flux on $\alpha$--$\beta$ plane.}
\begin{figure}[!htbp]
    \includegraphics[width=0.98\linewidth]{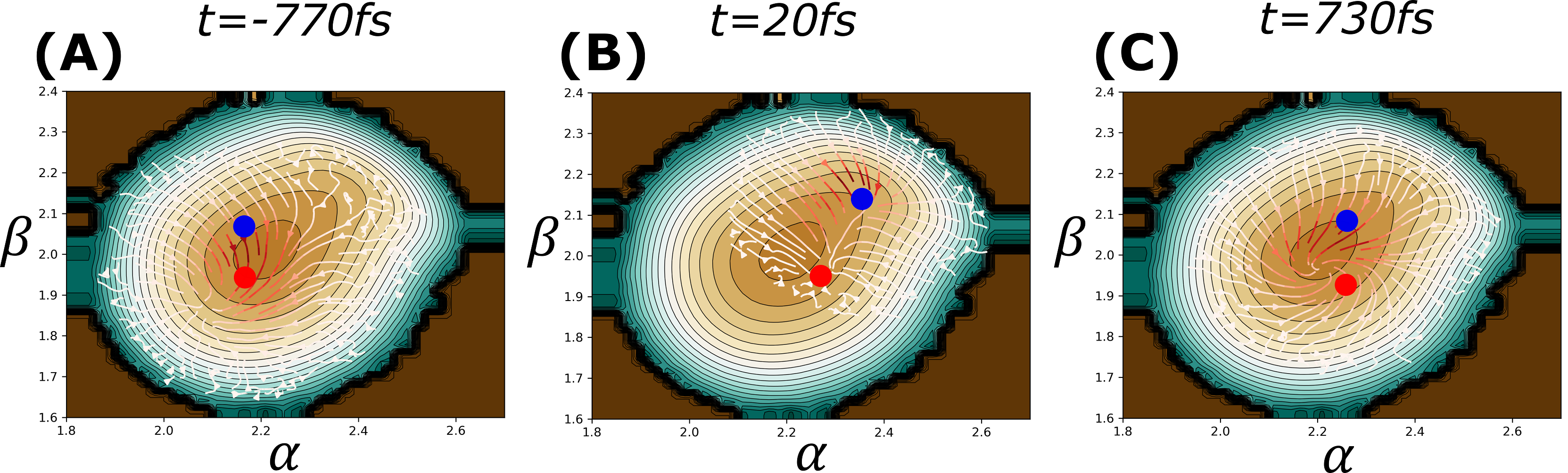}
    
    \caption{Flux on the $\alpha$--$\beta$ plane at three different time points.
    Blue dots are the locations of the probability peak at the
    current time, and red dots the peak location  $20$ fs afterwards.
    During the transition ($t=20$ fs), the region with strong flux is near the upper-right corner, and reverts back towards the original position after the transition. 
    }
    
    \label{fig:flux_Time_alpha_beta}
\end{figure}

We further examine the flux lines projected onto the $\alpha$--$\beta$ plane, and examine their relationship  with the topological changes of the probability peaks 
on the 3-d space of (time$, \, \alpha,\, \beta$).
Starting at $t=-770$ fs before the transition, the high flux regions (red arrow on Fig~\ref{fig:flux_Time_alpha_beta}A, and supplementary movie 5) move inside a small rectangle of $2.10<\alpha<2.30$, and $1.90<\beta<2.10$. 
At $t= 20$ fs near the transition time, the high flux  regions (red)  jump to the location of  $\alpha \approx 2.35$ and $\beta \approx 2.15$ (Fig~\ref{fig:flux_Time_alpha_beta}B). 
After the transition period at $t= 730$ fs, the high flux region  jumps back to the region where it was before the transition (Fig~\ref{fig:flux_Time_alpha_beta}C, and Supplementary movie 5). 

To examine the relationship between the fluxes and the time-dependent peak on the probability surface, we track the changes in the peak location at each time point (Fig~\ref{fig:flux_Time_alpha_beta}A-\ref{fig:flux_Time_alpha_beta}C). 
Here blue dots are the current location of the probability peak, and red dots are the location of peaks at about 20 $fs$ afterwards. %
It can be seen that that the probability peaks move along the flux direction over time in the $\alpha$--$\beta$ plane.


\newpage

\,

\newpage
\subsection{Birth location of the peaks}
\begin{table}[]
    \tiny
    \centering
    \begin{tabular}{lrrrrrrr}
\tiny
{} &  time &   phi &   psi &   theta1 &  Alpha &  Beta &   Birth Probability  \\
0  &     0 &  0.00 & -0.42 &     0.00 &   2.24 &  2.10 &             0.000459 \\
1  &  -145 & -1.26 &  0.01 &     0.39 &   2.18 &  1.94 &             0.000389 \\
2  &  -275 & -1.68 &  0.20 &    -0.18 &   2.13 &  1.94 &             0.000385 \\
3  &   -55 & -0.42 & -0.18 &     0.20 &   2.13 &  2.00 &             0.000372 \\
4  &   265 &  1.26 & -0.56 &     0.39 &   2.24 &  2.00 &             0.000363 \\
5  &   -95 & -0.84 &  0.01 &     0.39 &   2.18 &  2.00 &             0.000347 \\
6  &    55 &  0.42 & -0.18 &    -0.18 &   2.34 &  2.15 &             0.000311 \\
7  &    85 &  0.42 & -0.56 &     0.01 &   2.29 &  2.10 &             0.000307 \\
8  &   345 &  0.84 & -0.37 &     0.39 &   2.24 &  1.94 &             0.000301 \\
9  &   195 &  1.26 & -0.56 &     0.01 &   2.24 &  2.05 &             0.000299 \\
10 &   775 &  1.26 & -0.37 &     0.20 &   2.24 &  2.00 &             0.000233 \\
11 &   515 &  1.26 & -0.37 &    -0.37 &   2.24 &  2.05 &             0.000204 \\
12 &   465 &  1.26 & -0.37 &    -0.18 &   2.24 &  2.00 &             0.000203 \\
13 &   655 &  0.84 &  0.01 &     0.01 &   2.24 &  2.00 &             0.000181 \\
14 &   585 &  0.84 &  0.01 &    -0.18 &   2.24 &  2.10 &             0.000162 \\
15 &   555 &  0.84 &  0.01 &    -0.37 &   2.24 &  2.10 &             0.000161 \\
16 &   625 &  0.84 & -0.56 &     0.20 &   2.24 &  2.10 &             0.000146 \\
17 &  1195 &  0.84 & -0.56 &     0.39 &   2.24 &  2.05 &             0.000135 \\
18 &  1095 &  0.84 & -0.56 &     0.20 &   2.24 &  2.00 &             0.000130 \\
19 &  1275 &  1.26 & -0.37 &     0.01 &   2.24 &  2.05 &             0.000129 \\
20 &  1325 &  1.26 & -0.37 &    -0.18 &   2.24 &  2.05 &             0.000128 \\
21 &  1675 &  1.26 & -0.37 &     0.01 &   2.24 &  2.05 &             0.000112 \\
22 &  -695 & -1.26 &  0.96 &    -0.18 &   2.18 &  2.00 &             0.000105 \\
23 &  1475 &  0.84 &  0.01 &    -0.18 &   2.24 &  2.00 &             0.000105 \\
24 &  -705 & -1.26 &  0.01 &    -0.18 &   2.18 &  2.00 &             0.000099 \\
25 &  -545 & -0.84 &  0.77 &     0.20 &   2.13 &  2.00 &             0.000092 \\
26 & -1085 & -1.26 &  0.20 &    -0.18 &   2.18 &  2.00 &             0.000077 \\
27 &  -475 & -2.52 &  0.20 &     0.39 &   2.18 &  1.94 &             0.000076 \\
28 &  1995 &  0.84 & -0.56 &     0.20 &   2.24 &  2.00 &             0.000065 \\
29 & -1555 & -1.26 &  0.01 &     0.01 &   2.18 &  2.00 &             0.000044 \\
30 &  -955 & -2.94 &  1.34 &     0.01 &   2.18 &  1.94 &             0.000037 \\
31 & -1495 & -1.26 &  0.77 &    -0.18 &   2.18 &  1.94 &             0.000034 \\
32 & -1075 & -2.52 & -1.33 &    -0.18 &   2.18 &  1.94 &             0.000026 \\
33 & -1315 & -2.52 &  1.34 &     0.20 &   2.18 &  1.94 &             0.000025 \\
34 &  -795 & -1.26 & -1.33 &    -0.18 &   2.18 &  1.94 &             0.000022 \\
\end{tabular}
    \caption{Birth location of each peak and their birth probability}
    \label{tab:my_label}
\end{table}


\end{document}